\documentclass[twocolumn,superscriptaddress,aps,preprintnumbers,amsmath,amssymb,prd,nofootinbib %,dvipdfmx
]{revtex4-2}

\usepackage{graphicx}
\usepackage{epstopdf}
\usepackage{dcolumn}% Align table columns on decimal point
\usepackage{bm}% bold math
\usepackage{hyperref}
\usepackage{ulem}
\usepackage{color}
\usepackage{comment}

\usepackage{amsmath}

\def \e{{\rm e}}
\def\iu{{\rm i}}

\usepackage[dvipsnames]{xcolor}
\newcommand{\Red}[1]{\textcolor{red}{#1}}

\newcommand{\Mag}[1]{\textcolor{magenta}{#1}}
\renewcommand{\emph}[1]{\textit{#1}}

\newcommand{\red}[1]{\textcolor{red}{\textbf{#1}}}
\begin{document}

\title{
Axion dark matter search using arm cavity transmitted beams of gravitational wave detectors
}

\author{Koji Nagano}
\affiliation{Institute of Space and Astronautical Science, Japan Aerospace Exploration Agency, Sagamihara City 252-5210, Japan}
\author{Hiromasa Nakatsuka}
\affiliation{Institute for Cosmic Ray Research, University of Tokyo, Kashiwa 277-8582, Japan}
\author{Soichiro Morisaki}
\affiliation{Department of Physics, University of Wisconsin-Milwaukee, Milwaukee, WI 53201, USA}
\author{Tomohiro Fujita}
\affiliation{Waseda Institute for Advanced Study, Waseda University, Shinjuku, Tokyo 169-8050, Japan}
\affiliation{Research Center for the Early Universe,  The University of Tokyo, Bunkyo, Tokyo 113-0033, Japan}
\author{Yuta Michimura}
\affiliation{Department of Physics, University of Tokyo, Bunkyo, Tokyo 113-0033, Japan}
\affiliation{PRESTO, Japan Science and Technology Agency (JST), Kawaguchi, Saitama 332-0012, Japan}
\author{Ippei Obata}
\affiliation{Max-Planck-Institut f{\"u}r Astrophysik, Karl-Schwarzschild-Straße. 1, 85741 Garching, Germany}

\begin{abstract} % Length: (RPD) About 5% of article length & < 500 words

Axion is a promising candidate for ultralight dark matter which may cause a polarization rotation of laser light. Recently, a new idea of probing the axion dark matter by optical linear cavities used in the arms of gravitational wave detectors has been proposed [Phys. Rev. Lett. 123, 111301 (2019)]. In this article, a realistic scheme of the axion dark matter search with the arm cavity transmission ports is revisited. Since photons detected by the transmission ports travel in the cavity for odd-number of times, the effect of axion dark matter on their phases is not cancelled out and the sensitivity at low-mass range is significantly improved compared to the search using reflection ports. We also take into account the stochastic nature of the axion field and the availability of the two detection ports in the gravitational wave detectors. The sensitivity to the axion-photon coupling, $g_{a\gamma}$, of the ground-based gravitational wave detector, such as Advanced LIGO, with 1-year observation is estimated to be $g_{a\gamma} \sim 3\times10^{-12}$ GeV$^{-1}$ below the axion mass of $10^{-15}$ eV, which improves upon the limit achieved by the CERN Axion Solar Telescope.

% \RED{論文を書き始めるにあたっての内部向け要旨}\Red{ アクシオンは暗黒物質の候補。アクシオンは光共振器を使って探査する事ができる。本論文では、重力波検出器の腕共振器の透過光を使う手法を示す。重力波検出器は腕が長くハイパワーなのでショットノイズを小さくできる。さらに透過光を用いると、線形共振器で生じるパリティー変換による信号の打ち消しの影響を受けないので、低質量側での感度をよくできる。本論文では加えて、2つの検出器の相関解析や、アクシオン振動のコヒーレント時間が信号解析に与える影響についても定量的に示す。相関解析については、特に、重力波検出器の2つの透過光ポートを用いる方法について考える。コヒーレント時間は、本手法の特徴である低周波側の感度に影響するので考える。最終的な感度は、XXXで、YYY eV以下ではCASTを超える。}
\end{abstract}

%\keywords{inflation, primordial gravitational waves}
%\arxivnumber{1808.00548}

%\begin{flushright}

%\end{flushright}

\maketitle

%************************************************************************************%
%
%
%
%====================================================================================%
\section{Introduction}

Axion is a hypothetical pseudo-scalar field that was proposed by Peccei and Quinn to solve the strong CP problem in quantum chromo dynamics (QCD)~\cite{Peccei1977}. Their original idea is called ``QCD axion". %\NH{\sout{ in modern physics}}. 
Besides the QCD axion, string theory also predicts a plenty of axion-like particles via the compactifications of extra dimensions~\cite{Svrcek2006}. Typically, QCD axion and axion-like particles have two noteworthy features. First, their masses can be much lighter than 1 eV due to the shift symmetry. Second, they have an oscillatory feature leading to their behavior like a non-relativistic fluid in the Universe~\cite{Preskill:1982cy}. Owing to these two features, QCD axion or axion-like particles are considered to be a leading candidate for dark matter, among the ultra-light dark matter models~\cite{Marsh:2015xka,Ferreira:2020fam}. Hereafter in this article, we jointly call them ``axion."

To probe the axion, a variety of experiments and observations have been performed~\cite{Hagmann1998, CASTCollaboration2005, Vogel2013, Ehret2009, Betz2013, Tam2012, Sikivie2014, Kahn2016, Silva-Feaver2017, Brockway1996, Payez2015, Wouters2012, Marsh2017, Reynolds:2019uqt, Dessert:2020lil, Conlon2018, Aharonian2007, Conlon2013, Kohri:2017ljt, Moroi2018, Caputo2019, Gramolin:2020ict} (see also~\cite{Dias2014} for recent review).
They utilize a weak topological coupling between the axion and gauge bosons, such as photons, and detect an axion-photon conversion effect (dubbed ``Primakoff effect") under an external magnetic field~\cite{Sikivie1983}. For example, CERN Axion Solar Telescope (CAST) probes the axions produced in the Sun by converting the axion flux into X-rays with dipole magnets~\cite{CASTCollaboration2005}. 
Recently, some projects have also achieved almost the same limit as the CAST limit in a certain axion mass range below $\sim1$ neV~\cite{Gramolin:2020ict,Kahn2016}.
Another approach is the use of the astronomical telescopes to detect the electro-magnetic signals that is generated from the axion produced in the astronomical object, such as SN1987A. 
The axions produced in SN1987A can be converted to the gamma ray according to galactic magnetic field and the associated spectral modulation is potentially observable with the gamma-ray telescope~\cite{Payez2015}. 
Moreover, the recent observations of gravitational waves have provided the constraints independent from the above electromagnetic astrophysical observations
through the measurement of spinning black hole
~\cite{Tsukada:2018mbp}.
% \Mag{\sout{In addition to these \Blue{approaches}, there are many proposals to probe the axion via the Primakov effect, such as Search for Halo Axions with Ferromagnetic Toroids~\cite{Gramolin:2020ict}, A Broadband/Resonant Approach to Cosmic Axion Detection with an Amplifying B-field Ring Apparatus~\cite{Kahn2016}.
% \Blue{Some of them have been experimentally performed.}}}

Recently, a new search method for axion dark matter without using the Primakoff effect has been developed. The axion field coupled to photons behaves as a birefringent material in our universe by differentiating the phase velocities between left- and right-handed photons ~\cite{Carroll1990, Andrianov2010}.
Thanks to the recent development of the optics technology, many new approaches measuring the photon's birefringence caused by axion dark matter have been proposed~\cite{Melissinos2009, DeRocco2018, Obata2018, Liu2018, Nagano:2019rbw, Martynov:2019azm}.
% \Mag{\sout{These approaches utilize the modulation of the phase velocity of the speed of light caused by the oscillation of the axion field
%~\cite{Carroll1990, Andrianov2010}.
%}} 
These studies have led to an idea of using the existing or planned instruments originally for the different purposes, such as laser interferometric gravitational wave detectors~\cite{Aasi2015, Acernese2014, Somiya2012, Punturo2010, Abbott2017, Kawamura2008}. For example, there is a proposal to use long Fabry--Perot cavities in the gravitational-wave detectors for axion dark matter search~\cite{Nagano:2019rbw}. We call this scheme
\textit{ADAM-GD} (Axion DArk Matter search with Gravitational wave Detectors). %\YM{Agreed on this acronym?} 
ADAM-GD enables us to probe the axion dark matter with the mass less than $10^{-10}\text{ eV}$ that is a complement method to the axion dark matter search with the gravitational wave observation~\cite{Tsukada:2018mbp, Brito:2017wnc}.

In the ADAM-GD scheme, the axion dark matter can be searched for in the reflection and the transmission port of the Fabry--Perot cavity.
Of these two ports, the transmission port is more feasible than the reflection port in terms of the experimental setup since 
the reflection port is occupied by more main optics used for the gravitational wave detection, for example a signal extraction mirror~\cite{Meers1988}.
%Thus, as the first step for the experimental demonstration, more specific consideration is required. \YM{- omit this sentence?}
%\TF{[I agree to remove it]}

To demonstrate the benefits from the use of transmission port, in this article we develop a realistic performance of ADAM-GD with the transmission port and re-evaluate its potential sensitivity to the axion-photon coupling.
In particular, we revisit the response of the phase velocity modulation in the transmission port and consider the influence of the odd-number ways of optical path in the cavity on the axion signal, which was overlooked in the previous study~\cite{Nagano:2019rbw}.
To make the estimate more realistic, we take the randomness of the axion dark matter particles into account and introduce the associated stochastic behavior of axion field to the signal.
Moreover, we also perform the coherent analysis of two transmission ports of gravitational-wave detectors to improve the sensitivity level.
%\SM{To conservatively \Blue{(I don't feel that ``conservatively" certainly matches my thought. Do you have any idea? (NK))} \SM{(For a realistic estimate?)} estimate the sensitivity, we take into account the randomness of the axion field. \sout{For the consideration of the axion dark matter search in the ultralight mass range, we have to consider the dynamics of the axion field including stochastic oscillation}}.
%In addition, \Blue{the phase velocity modulation response of the transmission port in the gravitational-wave detectors is revisited.
%Specifically, we consider the effect of the odd-number optical path in the cavity that was not considered in~\cite{Nagano:2019rbw}.}
Note that, in this article, we set the natural unit $\hbar = c = 1$.

% \TF{[Don't we need ``This article is organized as ...''?]}
% \Blue{[How about this? (NK)]}
This article is organized as follows. Section~\ref{ss:Phase Velocity Modulation} describes a brief summary of the phase velocity variation caused by axion dark matter and the dynamics of axion field including its stochastic behavior. In section~\ref{ss:response}, the Fabry--Perot cavity response to the phase velocity modulation including the odd-number optical path effect is studied. Next section~\ref{Two detectors correlation} presents a study on the coherent analysis of two detection ports. Finally, section~\ref{sec_sensitivity} gives expected sensitivities of the gravitational wave detectors to the axion-photon coupling and discussion for the actual search.

%====================================================================================%
 
%%%%%%%%%%%%%%%%%%
\section{Phase Velocity Modulation} \label{ss:Phase Velocity Modulation}
%%%%%%%%%%%%%%%%%%

%The axion field follows a almost periodic oscillation with slowly changing random phase and amplitude due to a decoherence affect.
%Such decoherence and stochastic nature affects the determination of coupling constant.

In this section, we revisit the dynamics of axion field background and  compute its effective oscillation amplitude by including the stochastic effect which will be used to obtain the sensitivity of interferometer experiments.
%Although we assume the constant amplitude of axion field for the sensitivity of interferometer in Eq.~\eqref{eq_phi_simple}, the amplitude and phase fluctuates over the coherent time.
%The phase fluctuation weakens the sensitivity for long
%from $R^{-1/2}$ to $(\tau T)^{-1/4}$ as discussed in Sec.\ref{sec_sensitivity}.
%while the amplitude fluctuation affects the determination of the upper limit of coupling constant.
%In this section, we derive the prospected 95\% exclusion limit including the decoherent effect when the experiment find a null result.

%In this section, the dispersion relations of two circular-polarized photons modified in the axion field is explained.
The axion can couple to the photon through the Chern-Simons interaction,
\begin{align}
    \mathcal L_I = \frac{ g_{\text{a}\gamma} }{4} a(t) F_{\mu\nu} \tilde F^{\mu\nu},
\end{align}
where $a(t)$ is the axion field, $g_{\text{a}\gamma}$ is the axion-photon coupling constant, $F_{\mu\nu}$ is field strength of electromagnetic field and $\tilde F^{\mu\nu}\equiv \epsilon^{\mu\nu\rho\sigma}F_{\rho\sigma}/2$ is its Hodge dual with the Levi-Civita anti-symmetric tensor $\epsilon^{\mu\nu\rho\sigma}$. 
It is known that the background axion field  modifies the dispersion relation of photon.
Under the homogeneous axion field background, the left- (right-) circular polarized photon has a modified dispersion relation,
\begin{align}
    \omega^2_{\rm L/R} = k^2 (1\mp g_{\text{a}\gamma} \dot a /k), \label{eq:dispersion}
\end{align}
with its momentum $k$ and angular frequency $\omega_{\rm L/R}$.
Phase velocities of left-handed and right-handed photons are represented as
\begin{align}
    c_\text{L/R} &\equiv \frac{\omega_\text{L/R}}{k} \simeq 1 \mp \delta c(t) ,\\
    \delta c(t) &\equiv \frac{g_\mathrm{a\gamma} \dot a}{2k}.
    \label{eq_deltact}
\end{align} 
The difference of phase velocities between circular polarization modes rotates the direction of linear polarization of photon.
Therefore, the oscillating axion field converts a part of p-polarized light into s-polarized light or vice versa, which enables us to probe axion dark matter through gravitational-wave detectors, 
as discussed in the next section.

%Since the axion has a shift symmetry, the dispersion relation is modified only through its derivative, $\dot a$.
In the literature, %The dynamics of coherent 
the axion field is often modeled by a periodic oscillation, $a(t)=a_0 \cos(m t+\theta_0)$, with a constant amplitude $a_0$ as well as constant phase $\theta_0$, where $m$ is the mass of axion.
In reality, however, the axion field should be understood as a superposition of many classical waves with different velocities and phases~\cite{Foster:2017hbq}
\begin{align}
    a(t)
    &= 
    \sum_i \Lambda \cos\left(
    m(1+v_i^2 /2) t +\theta_i
    \right),
    \label{eq_a_sum}
\end{align}
where $\Lambda$, $v_i$ and $\theta_i$ are amplitude, velocity and phase of $i$-th wave, respectively.
The velocity dispersion of dark matter around the sun in our Galaxy is $v\sim 10^{-3}$.
Note that $\Lambda$ is properly normalized such that the expectation value of the energy density, $\rho_\mathrm{DM}=(\dot{a}^2+m^2 a^2)/2$,
reproduces the observed value in the local universe, $\rho_\mathrm{DM}\sim 0.4$ GeV/cm$^3$~\cite{deSalas:2020hbh}.
The oscillation frequencies of these waves are slightly different due to the velocity distribution, and the superposed axion $a(t)$ does not show a perfect coherent oscillation. Instead, the amplitude and phase of $a(t)$ varies in a so-called coherent time scale, $\tau\equiv 2\pi/(mv^2)$. Therefore,
after performing the summation of Eq.~\eqref{eq_a_sum}, the axion field is expressed as 
\begin{align}
    a(t)=a_0(t) \cos(m t+\theta_0(t)),
\end{align}
 where %the amplitude 
 $a_0(t)$ and %phase 
 $\theta_0(t)$ slowly change over %the time scale of
 $\tau$, while they can be approximated by constants for a shorter time scale than $\tau$.
%\Red{[From here, TF substantially revised the text.]}
Because of this time-evolving nature, $a_0(t)$ may significantly deviate from its mean value
\begin{align}
   \bar a \equiv \frac{\sqrt{2\rho_{\rm DM}}}{m}
\end{align}
when we conduct an experiment.
Therefore, the effective axion amplitude during experiments should be carefully considered to determine the sensitivities.
%In the rest of this section, we study the axion amplitude $a_0(t)$. The effect of the varying phase $\theta_0(t)$ will be discussed in Sec.~\ref{sec_sensitivity}.

To take into account the time evolving nature of $a_0$ and $\theta_0$, we approximate them by constants which stochastically change their values every coherent time $\tau$. 
We naively assume that the phase $\theta_0$ takes a value between $0$ and $2\pi$ in the equal probability.
The amplitude $a_0$ follows the Rayleigh distribution, because the sum of the amplitudes of the waves  with random phases in Eq.~\eqref{eq_a_sum} can be seen as the distance from the origin in a random walk process in the complex plane~\cite{Foster:2017hbq,Centers:2019dyn}.
%The sum of each modes can be approximated by the random walk in the complex plane, and $a_0$ measured at a certain moment behaves as the stochastic variable following the Rayleigh distribution~\cite{Foster:2017hbq,Centers:2019dyn} 
The probability distribution of $a_0$ is given by
\begin{align}
    P^{\rm (Ray)} (a_0) ~{\rm  d}a_0\equiv  \frac{2a_0}{\bar a^2} \exp\left( -\frac{a_0^2}{\bar a^2} \right)
     ~{\rm  d}a_0
    \label{eq_Rayleigh}.
\end{align}
%Here, $\bar a$ is the mean amplitude of the axion DM,
%\red{いままで$\rho_\text{DM}=$0.3 GeV/cm$^3$としていたが、最近の結果ではむしろ$\rho_\text{DM}=$0.4 GeV/cm$^3$の方がprefer。なので、感度計算とかは0.4を使ったほうが良いだろう。pointed by TF}
If the measurement time is shorter than the coherent time, $T_{\rm obs} \lesssim \tau$, only one realization of $a_0$ matters and Eq.~\eqref{eq_Rayleigh} suffices for our purpose.

For $T_{\rm obs} \gtrsim \tau$, the time variations of $a_0$ and $\theta_0$ become relevant.
%and $\theta(t)$ vary over the observation time.
%To avoid the complexity, 
We divide the measurement time into $N\equiv T_{\rm obs}/\tau$ domains in which $a_0$ and $\theta_0$ can be approximated by constants with different values.
%coherent patches with their duration of $\tau$. 
We call their values in the $i$-th domain $a_i$ and $\theta_i$.
Then, the axion field is approximately written as
\begin{equation}
a((i-1)\tau < t <i\tau)= a_i \cos(mt+\theta_i),
\end{equation}
where $a_i$ and $\theta_i$ are constant, and $i$ runs from 1 to $N$.
Using Eq.~\eqref{eq_deltact}, we find the phase modulation in the $i$-th domain as 
\begin{align}
    \delta c((i-1)\tau<t<i\tau)=\frac{g_{a\gamma}m a_i}{2k}\sin(mt+\theta_i)
\end{align}
In the data analysis, we work in the Fourier space and sum up not the amplitude but the power of the signal over the all domains. This is because the random phase $\theta_i$ prevents us from coherently adding up the amplitude. In each domain, the signal power is proportional to $g_{a\gamma}^2a_i^2$ and its sum yields $g_{a\gamma}^2\sum_ia_i^2$. Thus, the sensitivity of our experiment is characterized by the root mean square (RMS) of $a_i$, namely $A_{N} \equiv \sqrt{\sum_i a_i^2/N}$.
%For the later convenience, we calculate the probability distribution of the averaged variable, $a_{(N)}^2 \equiv N^{-1}\sum_i (a_i)^2$, where the $a_i$ is the realized amplitude at the $i$-th coherent patch.
From Eq.~\eqref{eq_Rayleigh}, we find the probability distribution of $A_N$ as (See appendix~\ref{app_derivation_PAN} for derivation.)
%\SM{(The second line in the equations below is not used for the derivation. Can we skip it?)} 
% \begin{align}
%   &P^{(N)}(A_N)
%     \nonumber
%     \\
%     &=
%     \int 
%     \left( 
%     \prod_i {\rm d} a_i
%     P^{\rm (Ray)} (a_i) 
%     \right)
%     \delta\left(  A_N - \sqrt{\frac{\sum_i a_i^2}{N}} \right)
%     \nonumber
% \\&=
%     \frac{2N^N}{\bar a\Gamma(N)}
%      \left(\frac{A_N}{\bar a}\right)^{2N-1}e^{-N(A_N/\bar a)^2}
%     ,
%     \label{eq_prob_N}
% \end{align}
\begin{equation}
  P^{(N)}(A_N) = 
    \frac{2N^N}{\bar a\Gamma(N)}
     \left(\frac{A_N}{\bar a}\right)^{2N-1}e^{-N(A_N/\bar a)^2}
    ,
    \label{eq_prob_N}
\end{equation}
where $\Gamma(n)$ is the gamma function.
Note that $P^{(1)}(a) = P^{\rm (Ray)} (a)$.
In the limit $N\to \infty$, $P^{(N)}(a) \to \delta(a - \bar a)$ and the stochastic effect vanishes.

\begin{figure}[htb]
  \centering
  \includegraphics[width=\columnwidth]{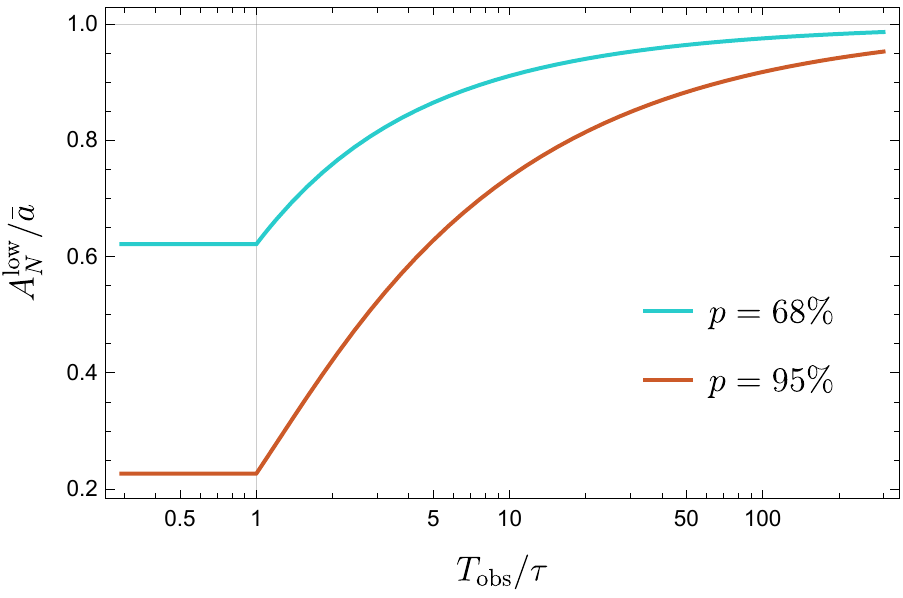}
  \caption{
  The probabilistic amplitude of axion field $A^{\mathrm{low}}_N(p)$ given in Eq.~\eqref{eq_def_a_alpha}. 
  We expect the actual axion RMS amplitude $A_N$ is larger than or equal to $A^{\mathrm{low}}_N(p)$ by
  probabilities of $p = 68\%$ (Cyan) and $95\%$ (Red).
  The horizontal axis denotes the experiment time divided by the coherent time, $T_\mathrm{obs}/\tau$, which corresponds to
  the number of time domain $N$ for $T_\mathrm{obs}/\tau> 1$,
  while $N$ remains unity for $T_\mathrm{obs}/\tau< 1$. 
  }
  \label{fig_a_alpha}
\end{figure}

Given that we basically measure $g_{a\gamma}^2A_N^2$, the sensitivity to $g_{\text{a}\gamma}$ would be weaker if $A_N$ happens to be smaller during our experiment.
To conservatively estimate the sensitivity by taking into account this stochastic effect,
we introduce a probabilistic RMS amplitude $A^{\mathrm{low}}_{N}(p)$ where $A_N\ge A^{\mathrm{low}}_{N}(p)$ occurs with probability of $p$. For instance, we expect that the actual RMS amplitude $A_N$ is larger than or equal to $A^{\mathrm{low}}_{N}(0.95)$ with a probability of 95\%. We compute $A^{\mathrm{low}}_{N}(p)$ as
\begin{align}
    p =
    \int^{\infty}_{A^{\mathrm{low}}_{N}(p)} {\rm d} a~
    P^{(N)} (a) 
    =
    \frac{\Gamma(N,N(A^{\mathrm{low}}_{N}(p)/\bar a)^2 )}{\Gamma(N)}
    ,
    \label{eq_def_a_alpha}
\end{align}
where $\Gamma(a,z)\equiv \int^\infty_z t^{a-1}e^{-t}{\rm d}t$ is the incomplete gamma function.
In Fig.~\ref{fig_a_alpha}, we show $A^{\mathrm{low}}_{N}(p)$ for $p =0.68$ and $0.95$.
%The sensitivity can decrease about factor $a_{\alpha}/\bar a$ for the unfortunate case with $1-\alpha$ probability.
For $T_{\rm obs}<\tau$, we measure only $N=1$ realization of the axion amplitude, and we find $A^{\mathrm{low}}_{1}(68\%) \approx 0.62\bar a$ and $A^{\mathrm{low}}_{1}(95\%) \approx 0.23\bar a$.
%For a longer observation time $T_{\rm obs}>\tau$, we solve Eq.~\eqref{eq_def_a_alpha} with $N = T_{\rm obs}/\tau$.
As the measurement time becomes longer, the stochastic effect becomes less important, since we average over $N$ realized values and $A^{\mathrm{low}}_N(p)$ converges into $\bar a$ in the limit $N\to \infty$.
%tends to have the sharp distribution around $a_{(N)}\sim \bar a$.
%As expected, the lower bound tends to converge in $\bar a$. 
%In appendix~\ref{app_derivation_PAN}, we derive simple fitting formulas as
We find fitting functions for $A^{\mathrm{low}}_{N}(68\%)$ and $A^{\mathrm{low}}_{N}(95\%)$ with respect to $N$ whose relative error is smaller than $1\%$,
\begin{align}
    \frac{A^{\mathrm{low}}_{N}(68\%)}{\bar a}
    &\simeq 
    \begin{cases}
          0.621
          & (N<1)
          \\
          1- 0.145N^{-0.956}-0.234 N^{-1/2}
          &  (1< N)
    \end{cases},
    \nonumber\\
    \frac{A^{\mathrm{low}}_{N}(95\%)}{\bar a}
    &\simeq 
    \begin{cases}
          0.226
          & (N<1)
          \\
          1+ 0.049N^{-3.410}-0.822 N^{-1/2}
          &  (1< N)
    \end{cases}.
    \label{eq_fitting_Alow}
\end{align}

%%%%%%%%%%%%%%%%%%%%%%%%%%%%
%%%%%%%%%%%%%%%%%%%%%%%%%%%%

%In short, we should measure the difference of phase velocities $\delta c_0$ to determine the coupling constant $g_{\text{a}\gamma}$ using Eq.~\eqref{eq_c0_g}.
Wrapping up the above arguments, we describe the way to find the sensitivity. 
In the experiment, we measure the oscillation amplitude $\delta c_0=g_{a\gamma}ma_0(t)/2k$ of 
the phase velocity difference in Eq.~\eqref{eq_deltact}.
We then process this data into $g_{a\gamma}m A_N/2k$ by dividing it into $N=T_\mathrm{obs}/\tau$ domains.
%Since Eq.~\eqref{eq_c0_g} depends on the amplitude of axion field $a_0(t)$, we calculate the expected lower bound of $a_0$ for 68\% and 95\% shown in Fig.~\ref{fig_a_alpha}. 
%Then, the coupling constant is given by
However, we cannot determine the axion RMS amplitude $A_N$ due to the stochastic effect.
Then we replace $A_N$ by $A^{\mathrm{low}}_N(p)$ and estimate the sensitivity based on
\begin{align}
    g_{a\gamma}
    =
    \frac{2k \delta c_0}{m A^{\mathrm{low}}_N(p)} . \label{eq:g_agamma}
\end{align}
The actual sensitivity to $g_{a\gamma}$ is better than or equal to the value estimated by the above equation by a probability of $p$.

Note that once we have performed the experiment and found null detection, we could put the upper bound of coupling constant using a more dedicated treatment~\cite{Centers:2019dyn}.

%We simply approximate the stochastic dynamics of field value by instant transition over each coherent time, where the $a$ has constant amplitude $a_i$ and phase $\theta_i$ at $t\in [i\tau,(i+1)\tau]$.
%$a_i$ and $\theta_i$ are the stochastic variables following the Rayleigh distribution in Eq.~\eqref{eq_Rayleigh} and uniform distribution over $\theta_i\in [0,2\pi]$.
%$a_i$ is the stochastic variable following the Rayleigh distribution in Eq.~\eqref{eq_Rayleigh}.
%\begin{align}
%	g_{{\rm a}\gamma}^{\rm (th)}
%    =
%	 s^{\rm (th)}
%     \bar a^{-1}
%    \frac{2k}{m}
%    \sqrt{\frac{S_{\rm shot}(m)}{T_{\rm obs}}  }
%    \quad
%    (T_{\rm obs}\lesssim \tau).
%    \label{eq_g_uppr_single}
%\end{align}
%\Red{レイリー分布での68\%, 95\% lineの定義。かつ測定時間ごとに定義する。}

\section{Axion search with an arm cavity transmitted beams of gravitational wave detectors}
\label{ss:response}

\begin{figure}[htb]
  \centering
  \includegraphics[width=\columnwidth]{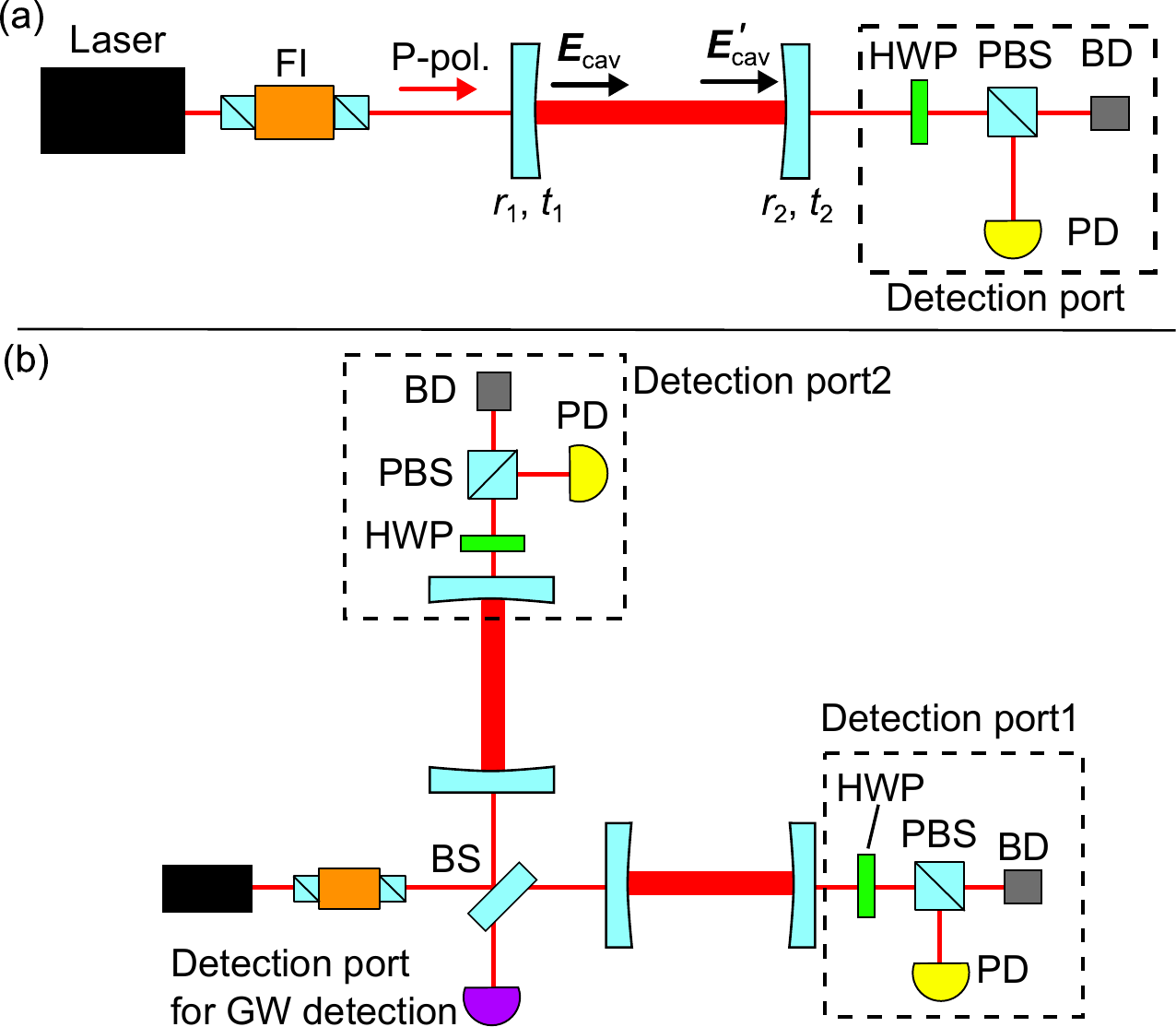}
  \caption{Schematic of experimental setup for axion search with (a) a linear optical cavity and (b) a gravitational wave detector. 
  FI, Faraday isolator; HWP, half wave plate; PBS, polarizing beam splitter; 
  PD, photodetector; BD, beam dump; BS, beam splitter. In the detection port at the cavity transmission, polarization analysis is performed with the HWP, PBS, and PD, and the axion signal can be detected. Components for phase measurement are not shown. Two PBSs in FI are placed 
  rotated by 45 degrees along the optical path.}
  \label{fig:Setup}
\end{figure}

In this section, we express the sensitivity of the transmission port of the gravitational wave detector. First, we consider the response function for the transmission port. 
The setup is shown in figure \ref{fig:Setup}-(a). The laser light with a wavelength $\lambda = 2\pi/k$ is injected to the optical cavity in purely p-polarized light. As explained in the previous section, the axion field generates s-polarized signal light from the p-polarized light. The optical cavity is composed of the input and the end mirror. The input (end) mirror has reflectivity and transmissivity of $(r_1, t_1)$ ($(r_2, t_2)$). Hereafter, we assume the optical cavity meets the resonant condition. In other words, the cavity length, $L_\mathrm{cav}$, is kept to be integer multiple of the half wavelength of the laser light with, for example, Pound--Drever--Hall technique~\cite{Drever1983}. In this condition, the laser light including the s-polarized light of the axion signal is enhanced in the cavity as explained later. The enhanced signal is observed in the detection port at the cavity transmission. The detection port consists of the light polarization detector with a half-wave plate (HWP) and a polarizing beam splitter (PBS).

Here, we explain how the laser light is enhanced in the cavity. 
The input laser light is expressed as
\begin{equation}
    \bm{E}_\mathrm{in}(t) = \bm{E}^\mathrm{p}(t) = E_0 e^{ikt} (\bm{e}^L \ \bm{e}^R)\frac{1}{\sqrt{2}}\begin{pmatrix}
      1 \\
      1
\end{pmatrix},
\end{equation}
where $\bm{E}^\mathrm{p}(t)$ is the electric vector of the p-polarized light, and $\bm{e}^L$ and $\bm{e}^R$ are basis vectors of the left- and right-handed photon, respectively. When the cavity is in the resonant condition, the electric field of the input laser light is enhanced by $\frac{1}{1-r_1 r_2}$~\cite{Yariv2007}. In the presence of the background axion field, the left- and right-handed light has the phase variation with opposite sign. Therefore, the intra-cavity electric field at the input mirror, $\bm{E}_\text{cav}(t)$, is denoted as~\cite{Nagano:2019rbw},
\begin{align}
\bm{E}_\text{cav}(t) =& \frac{t_1 E_0e^{i k t}}{1-r_1r_2}
\begin{pmatrix}
        \bm{e}^\text{L} & \bm{e}^\text{R}
\end{pmatrix} \nonumber \\
& \ \times
\begin{pmatrix}
      1 + i \delta \phi(t) & 0 \\
      0 & 1 - i \delta \phi(t)
\end{pmatrix}
\frac{1}{\sqrt{2}}
\begin{pmatrix}
      1 \\
      1
\end{pmatrix} \\
=& \frac{t_1 }{1-r_1r_2}
\left[\bm{E}^\text{p}(t) 
- \delta \phi(t)\bm{E}^\text{s}(t)\right],
\end{align}
where 
$\bm{E}^\mathrm{s}(t)$ is the electric vector of the s-polarized light represented as
\begin{equation}
  \bm{E}^\mathrm{s}(t) = E_0 e^{ikt} (\bm{e}^L \ \bm{e}^R)\frac{1}{\sqrt{2}i}\begin{pmatrix}
      1 \\
      -1
\end{pmatrix},
\end{equation}
and $\delta \phi(t)$ is the polarization angle rotated by the phase velocity deference $\delta c$,
\begin{align}
 \delta \phi(t) &\equiv \int^\infty_{-\infty} 
 \tilde{\delta c}(m) H_\text{a}(m) e^{im t} \frac{dm}{2\pi}.
\end{align}
Here, $\tilde{\delta c}(m)$ is the Fourier transformation of $\delta c(t)$, $\delta c(t) \equiv \int^\infty_{-\infty}\tilde{\delta c}(m)e^{imt} \frac{dm}{2\pi}$ and $H_\text{a}(m)$ is a response function of cavity,
\begin{equation}
 H_\text{a}(m) \equiv i \frac{k}{m} 
 \frac{4 r_1r_2\sin^2\left(\frac{m L_\mathrm{cav}}{2}\right)}
 {1 - r_1 r_2 e^{-i2m L_\mathrm{cav}}}  \left(-e^{-imL_\mathrm{cav}}\right) \label{eq:Ha},
\end{equation}
which corresponds to the response at the reflection port (the detection port nearby the front mirror)~\cite{Nagano:2019rbw}.

The intra-cavity electric field at the end mirror, $\bm{E}^\prime_\text{cav}(t)$, is obtained by applying the transfer matrix for one way translation, $T(t)$, to that at the input mirror. 
$\bm{E}^\prime_\text{cav}(t)$ is represented as
\begin{align}
\bm{E}^\prime_\text{cav}(t+L_\mathrm{cav}) =& \frac{t_1 E_0e^{i k t}}{1-r_1r_2}
\begin{pmatrix}
        \bm{e}^\text{L} & \bm{e}^\text{R}
\end{pmatrix} 
T(t+L_\mathrm{cav}) \nonumber \\
& \ \times
\begin{pmatrix}
      1 + i \delta \phi(t) & 0 \\
      0 & 1 - i \delta \phi(t)
\end{pmatrix}
\frac{1}{\sqrt{2}}
\begin{pmatrix}
      1 \\
      1
\end{pmatrix} \label{eq:Ecavp}
\end{align} 
where 
\begin{equation}
T(t) \equiv 
\begin{pmatrix}
        e^{-i\phi^\text{L}(t)} & 0 \\
        0 & e^{-i\phi^\text{R}(t)}
\end{pmatrix}, \label{eq:T}
\end{equation}
\begin{equation}
\phi^\text{\text{L/R}}(t) \equiv k L_\mathrm{cav} \mp  k \int^t_{t-L_\mathrm{cav}} \delta c(t^\prime) dt^\prime . \label{eq:deltaphip}
\end{equation}
Notice that, since it takes $L_\mathrm{cav}$ for the electric field to travel from the input mirror to the end mirror, we have to consider $\bm{E}^\prime_\text{cav}(t+L_\mathrm{cav})$ if we start from $\bm{E}_\text{cav}(t)$.
When we assume $|\delta \phi^\prime(t)| \ll 1$ and $2k L_\mathrm{cav} = 2 \pi l \ \ (l\in\mathbb{N})$ (resonant condition of the cavity), Eq.~(\ref{eq:T}) can be deformed as
\begin{equation}
    T(t) \simeq (-1)^l
    \begin{pmatrix}
        1 + i\delta \phi^\prime(t) & 0 \\
        0 & 1 - i\delta \phi^\prime(t)
    \end{pmatrix}
\end{equation}
Consequently, Eq.~\eqref{eq:Ecavp} is denoted as
\begin{widetext}
\begin{align}
\bm{E}^\prime_\text{cav}(t+L) &\simeq (-1)^l\frac{t_1 E_0e^{i k t}}{1-r_1r_2}
\begin{pmatrix}
        \bm{e}^\text{L} & \bm{e}^\text{R}
\end{pmatrix} 
\begin{pmatrix}
      1 + i (\delta \phi(t) + \delta \phi^\prime(t+L_\mathrm{cav})) & 0 \\
      0 & 1 - i (\delta \phi(t) + \delta \phi^\prime(t+L_\mathrm{cav}))
\end{pmatrix}
\frac{1}{\sqrt{2}}
\begin{pmatrix}
      1 \\
      1
\end{pmatrix}, \label{eq:Ecavp2} \\
&= (-1)^l\frac{t_1}{1-r_1r_2}\left[\bm{E}^\text{p}(t) 
- (\delta \phi(t)+\delta \phi^\prime(t+L_\mathrm{cav}))\bm{E}^\text{s}(t)\right]. \label{eq:Ecavp2-1}
\end{align} 
\end{widetext}
Here, the second or higher order terms of $\delta \phi(t)$ and $\delta \phi^\prime(t)$ are ignored. 

\begin{figure}[htb]
  \centering
  \includegraphics[width=\columnwidth]{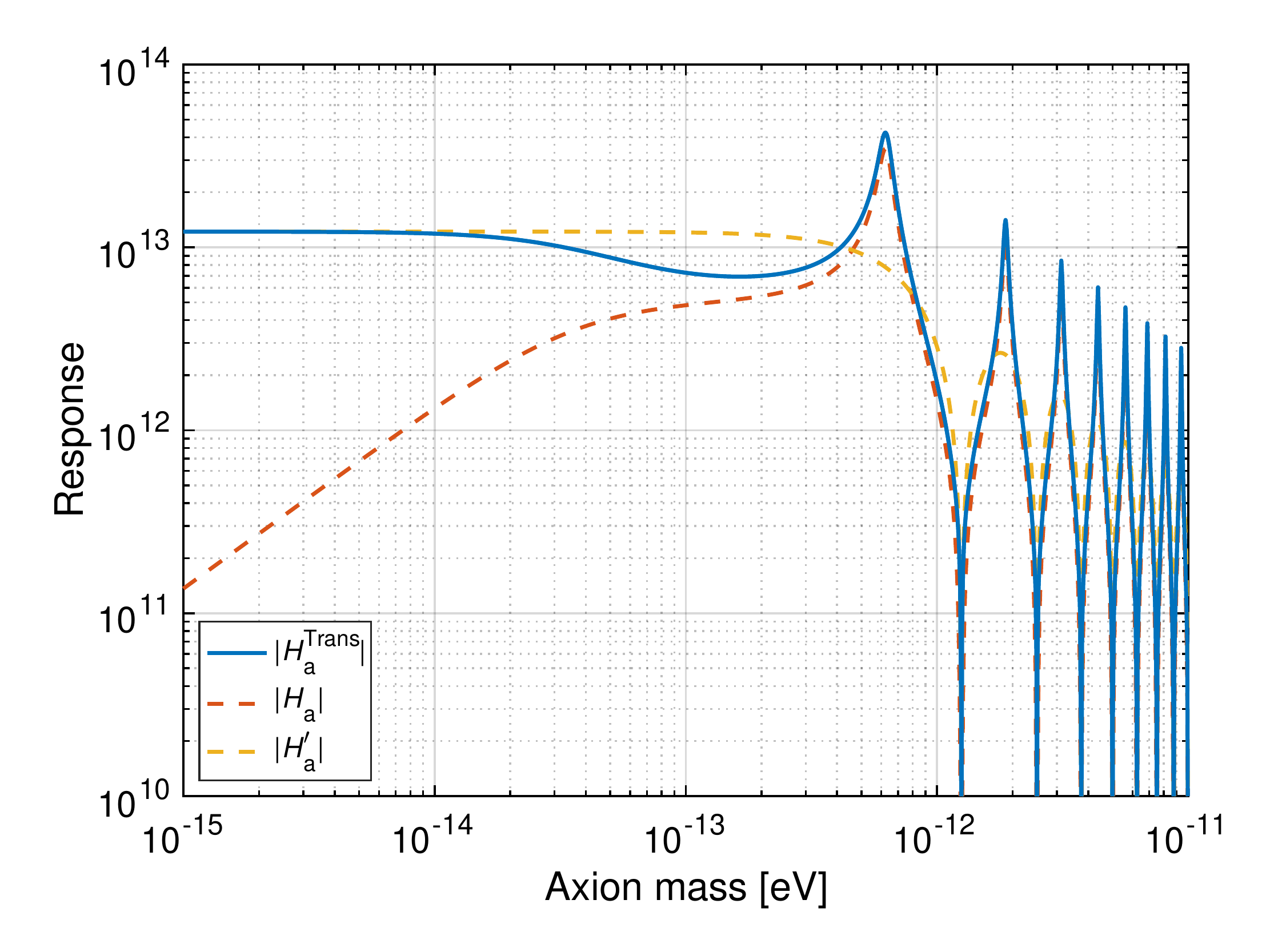}
  \caption{Response functions for the transmission port $H^\text{Trans}$ (blue) of DECIGO. $H^\text{Trans}$ has two contributions, $H_a(m)$ (red dashed) and $H_a'$ (yellow dashed) given in Eqs.~\eqref{eq:Ha} and \eqref{eq:Ha prime}, respectively.}
  \label{fig:Response}
\end{figure}

From Eq.~(\ref{eq:deltaphip}), $\delta \phi^\prime(t)$ is expressed as
\begin{align}
    \delta \phi^\prime(t+L_\mathrm{cav}) 
    &= k \int^{t+L_\mathrm{cav}}_{t} \delta c(t^\prime) dt^\prime \label{eq:Ecavp3}\\
    &= k \int^{t+L_\mathrm{cav}}_{t} \int^\infty_{-\infty} \tilde{\delta c}(m) e^{imt^\prime}\frac{dm}{2 \pi} dt^\prime \label{eq:Ecavp4} \\
    &\equiv \int^\infty_{-\infty}\tilde{\delta c}(m) H^\prime_\text{a}(m) e^{imt} \frac{dm}{2\pi}, \\
    H^\prime_\text{a}(m) &= \frac{2k}{m}e^{i\frac{mL_\mathrm{cav}}{2}}\sin\left(\frac{mL_\mathrm{cav}}{2}\right).
    \label{eq:Ha prime}
\end{align}
%\TF{\sout{From equation~ (\ref{eq:Ecavp3}) to equation (\ref{eq:Ecavp4}), Fourier transformation, $\delta c(t) \equiv \int^\infty_{-\infty}\tilde{\delta c}(m)e^{imt} \frac{dm}{2\pi}$, is performed.}} 
As a result, a total response function for the trasnmission port is defined as
\begin{equation}
    H^\text{Trans}_\text{a}(m) \equiv H_\text{a}(m) + H^\prime_\text{a}(m). \label{eq:Htrans}
\end{equation}
Figure~\ref{fig:Response} shows the absolute value of the response function of the transmission and reflection port. Here, we use the parameter set of DECIGO shown in table~\ref{tab:ITFparameters}.
As has been found in the previous work~\cite{Nagano:2019rbw}, the response at the reflection port $H_\text{a}(m)$ degrades the sensitivity for the low mass range because the phase modulation is cancelled out 
between the %on
going and returning ways of optical path in the cavity.
This %unlucky 
situation is caused by the parity transformation at the reflection of mirrors.
On the other hand, remarkably, the transmission port exhibits a sizable response for the low mass range.
This is because the cancellation of the phase velocity modulation does not occur due to the odd-number ways of optical path in the cavity.
Therefore, the use of transmission port is highly advantageous in comparison with that of reflection port for the low-mass axion searches.

\begin{table*}[htb]
\centering
 \caption{Parameters of considered gravitational wave detectors. Note that $P_0$ is the input beam power to the cavity enhanced by the power recycling cavity for the KAGRA-like, the aLIGO-like, and the CE-like detector~\cite{Drever1983a}. %\YM{Let's add CE!}
 }
 \label{tab:ITFparameters}
 \begin{tabular}{c|c|c|c|c}
  Similar detector      &       $L_\mathrm{cav}$     [m] &   $P_0$ [W]       &       $\lambda$ [$\times 10^{-9}$ m]    &       $(t_1^2, t_2^2)$ [ppm]  
  \\ \hline
 % CE~\cite{Abbott2017} &       $4\times10^4$   &       600     &       1550    &       ($1.2\times10^3$, 5)      \\ \hline
  KAGRA~\cite{Somiya2012}        &       $3\times10^3$   &       335    &       1064    &       ($4\times10^3$, 7)\\ \hline
  aLIGO~\cite{Aasi2015}        &       $4\times10^3$   &       2600    &       1064    &       ($1.4\times10^4$, 5)\\ \hline
  CE~\cite{Abbott2017}       &       $4\times10^4$   &       600    &       1550    &       ($1.2\times10^3$, 5)  \\ \hline 
  DECIGO~\cite{Kawamura2008}    &       $10^6$  &       5       &       515     &       ($3.1 \times 10^5$, $3.1 \times 10^5$)       
 \end{tabular}
\end{table*}

\section{Coherent analysis of two detection ports}
\label{Two detectors correlation}

%\SM{(Notations changed. More explanations added)} 

For detectors and mass range we consider in this study, the coherent length of the axion field $v \tau \sim 10^6 \text{ m}\times (10^{-9} \text{ eV}/m)$ is larger than the sizes of detectors. In this case, the two detection ports of a detector observe the signal with common phase.
% The two detection ports of an interferometeric gravitational-wave detector observe the signal with common phase.
% \Red{
% Here, we assume that the typical size of the detector is nor larger than the coherent length of the axion field oscillation represented as $v \tau \sim 10^6 \text{ m}\times (10^{-9} \text{ eV}/m)$.
% }
We show that coherently analysing them improves the sensitivity.
If $T_\text{obs} \ll \tau$, the signal is concentrated in a single frequency bin.
The Fourier component of data from the $i$--th port at that frequency bin  is given by
\begin{equation}
d_i = s + n_i~~~(i=1,2),
\end{equation}
where $s$ is the axion signal and $n_i$ is the noise of the $i$--th port.
If we only had the first port, we would use the following single-port detection statistic,
\begin{equation}
\rho_1 \equiv |d_1|^2.
\end{equation}
In the absence of signal, the probability distribution of $\rho_1$ is given by
\begin{equation}
p(\rho_1) = \frac{1}{2 S} e^{-\frac{\rho_1}{2 S}} \label{eq:rho1_pdf}
\end{equation}
where we assume that the noise is stationary and Gaussian, and $S \equiv \left<(\Re n_1)^2\right> = \left<(\Im n_1)^2\right>$.

Since we have two ports observing the same signal, we consider the coherent sum as detection statistic,
\begin{equation}
\rho \equiv |d_1 + d_2|^2. \label{eq:coherent_snr}
\end{equation}
The contributions from the signal to $\rho$ are $4$ times larger than those to $\rho_1$.
In the absence of signal, $\rho$ follows the same distribution as \eqref{eq:rho1_pdf} with $S$ replaced by $2(S + C)$, where we assume the variances of $n_1$ and $n_2$ are the same, and $C \equiv \left<\Re n_1 \Re n_2\right> = \left<\Im n_1 \Im n_2\right>$.
It means the noise contributions to $\rho$ are larger than those to $\rho_1$ by a factor of $2(1 + C / S)$.
Thus, the signal-to-noise ratio (SNR) becomes larger by a factor of $2 / (1 + C / S)$.
Since $\rho_1$ and $\rho$ are proportional to the square of the amplitude of signal, the constraints on the amplitude are improved by a factor of $\sqrt{2 / (1 + C / S)}$.
Hence, it is improved by $\sqrt{2}$ in the uncorrelated case, $C=0$.

If $T_\text{obs} \gtrsim \tau$, the sum of detection statistic over multiple frequency bins or segments whose durations are shorter than $\tau$ needs to be calculated~\cite{Morisaki:2018htj}.
Even in this case, we can compute the coherent detection statistic instead of the single-port one for each frequency bin or segment, and take their sum.
It leads to an improvement in the sensitivity by the same factor as that for $T_\text{obs} \ll \tau$.
In conclusion, the constraints on the amplitude, or the coupling constant, are improved by a factor of $\sqrt{2 / (1 + C / S)}$ thanks to the coherent analysis.

%%%%%%%%%%%%%%%%%
\section{Experimental sensitivity}
\label{sec_sensitivity}
%%%%%%%%%%%%%%%%

Based on the discussions in the previous sections, we find
%\TF{[`refind' isn't on dictionaries. Let's say `find' as usual.]}
the sensitivity to the axion-photon coupling in ADAM-GD scheme of transmission port. We assume that the primary source of noise is the quantum shot noise. Other technical noises are discussed later. First, the sensitivity of the one transmission port is considered as shown in figure~\ref{fig:Setup}-(a). We can estimate the shot noise for the axion dark matter search by estimating the amount of vacuum fluctuation of electric field in the photo detector~\cite{Kimble2001a}. Specifically, the shot noise spectrum, $S_\mathrm{shot}$, equivalent to $\tilde{\delta c}(m)$ is estimated from the ratio of the s-polarized field generated by axion dark matter to the vacuum fluctuation in the cavity transmission. Using equation (\ref{eq:Ecavp2-1}), $S_\mathrm{shot}$ is denoted as
\begin{align}
    \sqrt{S_\mathrm{shot}(m)} =& \frac{1}{\frac{t_1t_2}{1-r_1r_2}H^\text{Trans}_\text{a}(m)E_0} \\
    =& \frac{1}{\frac{t_1t_2}{1-r_1r_2}H^\text{Trans}_\text{a}(m)\sqrt{\frac{2P_0}{k}}}, \label{eq:shotnoise}
\end{align}
where $P_0$ is the power incident to the cavity. Here, we have used the fact that the spectrum of the vacuum fluctuation is unity. Note that the dimension of the electric field is [$\sqrt{\text{Hz}}$] in common with~\cite{Kimble2001a}.

The SNR of $\delta c_0$ depends on whether the observation time is longer than the axion oscillation coherent time or not as~\cite{Budker:2013hfa}
\begin{equation}
    \text{SNR} = 
    \begin{cases}
      \frac{\sqrt{T_\mathrm{obs}}}{2 \sqrt{S_\mathrm{shot}(m)}} \delta c_0 \ \ \ (T_\mathrm{obs}  \lesssim \tau) \\
      \frac{(T_\mathrm{obs} \tau)^{1/4}}{2 \sqrt{S_\mathrm{shot}(m)}} \delta c_0 \ \ \ (T_\mathrm{obs} \gtrsim \tau)
    \end{cases}.
\end{equation}
If we set detection threshold as the unity SNR, we can denote the detection limit of $\delta c_0$ as
\begin{equation}
\delta c_0 \simeq 
  \begin{cases}
    \frac{2}{\sqrt{T_\text{obs}}} \sqrt{S_\text{shot}(m)} & (T_\text{obs} \lesssim \tau) \\
    \frac{2}{(T_\text{obs} \tau)^{1/4}}\sqrt{S_\text{shot}(m)} & (T_\text{obs} \gtrsim \tau)
  \end{cases} .
  \label{eq_deltac_T_tauT}
\end{equation}
% \red{$\delta c_0$と$a_0$?の関係式($\delta c_0 = \frac{g_{a\gamma}a_0 m}{2k}$)に対応するものをどこかで書きたい。}
Considering the stochastic effect shown in Eq.~(\ref{eq:g_agamma}) and figure~\ref{fig_a_alpha}, the sensitivity to $g_\mathrm{a\gamma}$ of one cavity is expressed as %to be $a_{68\%} = 0.39\bar a$
% \begin{equation}
% \delta c_0 \simeq 
%   \begin{cases}
%     0.39\frac{2}{\sqrt{T_\text{obs}}} \sqrt{S_\text{shot}(m)} & (T_\text{obs} \lesssim \tau) \\
%     \frac{\bar a}{a_\text{68\%}(T_\text{obs}/\tau)}\frac{2}{(T_\text{obs} \tau)^{1/4}}\sqrt{S_\text{shot}(m)} & (T_\text{obs} \gtrsim \tau)
%   \end{cases} .
%   \label{eq_deltac_T_tauT}
% \end{equation}
% The factor of 0.39 for $T_\text{obs} \lesssim \tau$ comes from $a_\text{68\%} = 0.39\bar a$.
% Consequently, the sensitivity to $g_\mathrm{a\gamma}$ is expressed as
\begin{align}
g_{\text{a}\gamma}(m) &\simeq 1.9\times10^{12} \ \mathrm{GeV}^{-1} 
\left( \frac{1064 \text{ nm}}{\lambda} \right) 
\notag\\&\quad\times \frac{\bar a}{A^{\mathrm{low}}_N(68\%)}
  \begin{cases}
    \sqrt{\frac{S_\text{shot}(m)}{T_\text{obs}}}  
    & (T_\text{obs} \lesssim \tau) \\
    \frac{\sqrt{S_\text{shot}(m)}}{(T_\text{obs} \tau)^{1/4}} 
    & (T_\text{obs} \gtrsim \tau)
  \end{cases} .
\end{align}
% \red{ここ1064 nmにして$\rho_\mathrm{DM} = 0.4$ GeV/cm$^3$にした。}
Note that the unity SNR corresponds to $p = 0.68$.
% Note that how the sensitivity is improved according to the observation duration, $T_\mathrm{obs}$, depends on $T_\mathrm{obs}$ is larger than the axion coherent oscillation time, $\tau$.

Since the shot noise of the two detection ports are uncorrelated, the sensitivity to $g_\mathrm{a\gamma}$ can be improved by a factor of $\sqrt{2}$ using the coherent analysis shown in Section~\ref{Two detectors correlation}. Thus, the sensitivity to $g_\mathrm{a\gamma}$ of the gravitational wave detectors is represented as
\begin{align}
g_{\text{a}\gamma}(m) &\simeq 1.9\times10^{12} \ \mathrm{GeV}^{-1} 
\left( \frac{1064 \text{ nm}}{\lambda} \right) 
\notag\\&\quad\times \frac{\bar a}{\sqrt{2}A^{\mathrm{low}}_N(68\%)}
  \begin{cases}
    \sqrt{\frac{S_\text{shot}(m)}{T_\text{obs}}}  
    & (T_\text{obs} \lesssim \tau) \\
    \frac{\sqrt{S_\text{shot}(m)}}{(T_\text{obs} \tau)^{1/4}} 
    & (T_\text{obs} \gtrsim \tau)
  \end{cases} .
\end{align}

\begin{figure}[htb]
  \centering
  \includegraphics[width=\columnwidth]{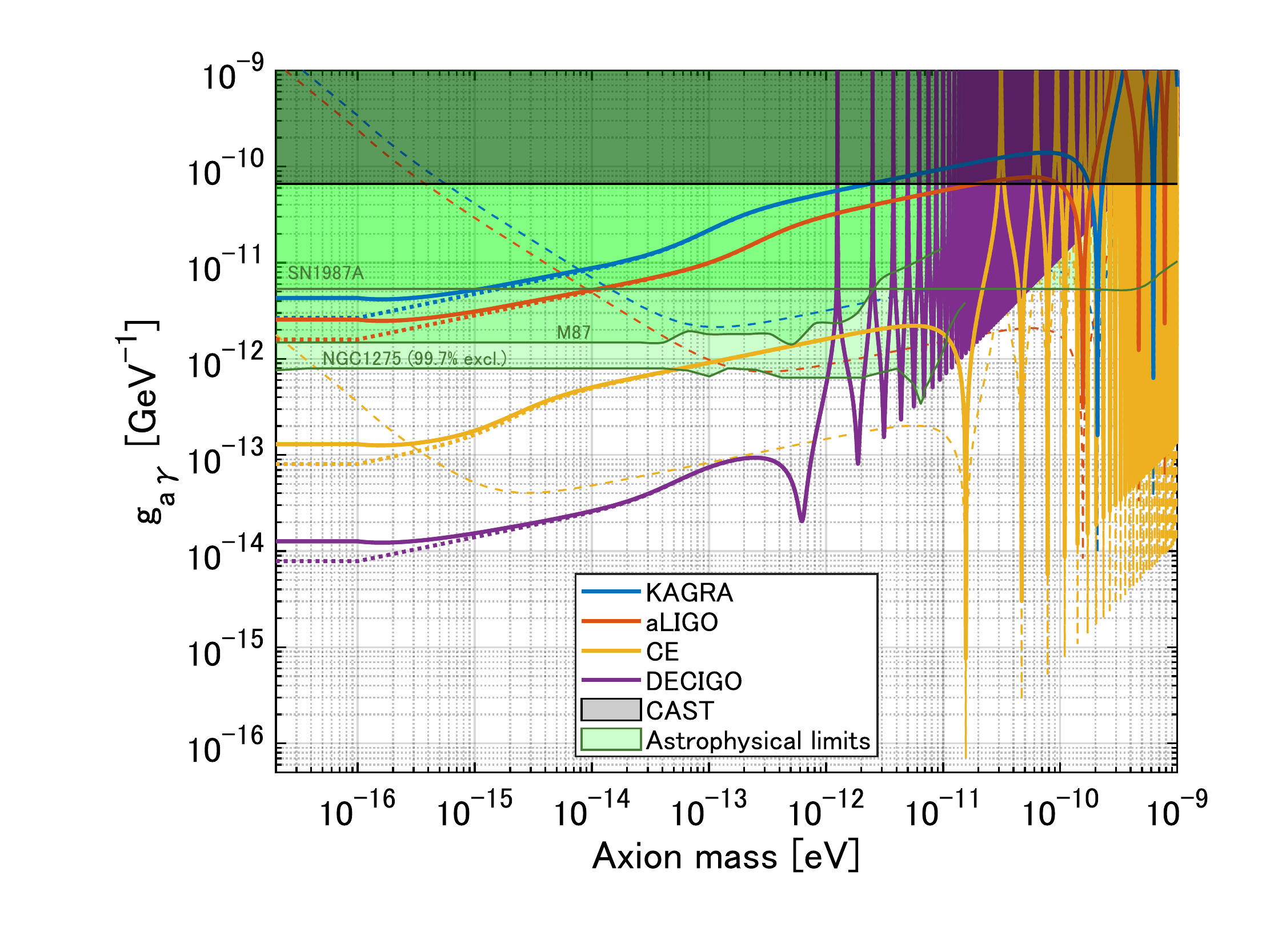}
  \caption{Sensitivity comparison of the several parameter sets shown in table 
  \ref{tab:ITFparameters}.
  The dotted lines represent the sensitivity without modification of the stochastic effect.
  Although the higher mass range seems to be filled, 
  they have sensitivity peaks at mass of $m=\pi(2N-1)/L_\mathrm{cav} \ (N\in\mathbb{N})$.
  The gray and green band express the current limits provided by 
  CAST~\cite{CASTCollaboration2005} and the astrophysical observations (SN1987A~\cite{Payez2015}, M87~\cite{Marsh2017}, and NGC 1275~\cite{Reynolds:2019uqt}). 
  The dashed lines are sensitivities using reflection port shown in~\cite{Nagano:2019rbw} with modification of the stochastic effect.% \red{Cite Nature Physics volume 17, pages79–84(2021) somewhere}\red{いままで$\rho_\text{DM}=$0.3 GeV/cm$^3$としていたが、最近の結果ではむしろ$\rho_\text{DM}=$0.4 GeV/cm$^3$の方がprefer。なので、感度計算とかは0.4を使ったほうが良いだろう。pointed by TF}
  % Numerical calculation of Inverse Gamma function from eq. (12) is imported. 2021/05/12 KN
  }
  \label{fig:sensitivity1}
\end{figure}

Figure \ref{fig:sensitivity1} presents the estimated sensitivity to $g_\mathrm{a\gamma}$ of the gravitational wave detectors as shown in figure~\ref{fig:Setup}-(b) with the parameter sets given in table~\ref{tab:ITFparameters}. 
We consider similar parameters to the existing or planned detectors, KAGRA~\cite{Somiya2012}, Advanced LIGO (aLIGO)~\cite{Aasi2015}, Cosmic Explorer (CE)~\cite{Abbott2017}, and DECIGO~\cite{Kawamura2008}.
Here, we assume $1$-year observation and axion is a dominant component of dark matter. 
The kink around $10^{-16}$ eV %is due to the coherence time effect of the axion oscillation.
is due to the phase velocity modulation of the axion oscillation.
With the DECIGO-like detector, the sensitivity is estimated to be $g_\mathrm{a\gamma} \simeq 1.3\times 10^{-14}$ GeV$^{-1}$ around $m<10^{-16}$ eV.
%We are planning to start the observation with KAGRA in next year. 
The sensitivity of the KAGRA-like (aLIGO-like) detector with transmission port overcomes the CAST limit in mass range less than $2\times 10^{-12}$ ($2\times 10^{-11}$) eV.
The sensitivity of the CE-like detector is better than $g_\mathrm{a\gamma} = 2\times 10^{-13}$ GeV$^{-1}$ in the mass range less than $10^{-15}$ eV.
In addition, figure \ref{fig:sensitivity1} shows that the observation with the transmission port in KAGRA-like, aLIGO-like, and CE-like detectors complements that with the reflection port shown in~\cite{Nagano:2019rbw}.
Figure~\ref{fig:sensitivity1} indicates that the sensitivity to $g_\mathrm{a\gamma}$ of the ground-based detectors are almost comparable to the  observation of SN1987A~\cite{Payez2015} below $\sim10^{-15}$-$10^{-14}$ eV and a bit weaker than the limit from the observations of M87~\cite{Marsh2017} and NGC 1275~\cite{Reynolds:2019uqt} even below $10^{-16}$ eV. %\YM{\sout{In spite of that, the observation with the ground-based detectors are still valuable. This is because our proposal does not need astronomical model assumption. Therefore the observation with the ground-based detectors are complementary to the astronomical observation.} (Sounds redundant to me) 
The observations with the ground-based detectors are complementary to the astronomical observations since our method is independent of astronomical model assumptions.
% \red{天文観測を超えないのでそのexcuse。} 

We note that figure~\ref{fig:sensitivity1} shows the shot-noise-limited sensitivities, and various practical noises should be further investigated. For example, the polarization rotation of the input laser is the possible noise source although the effect of the noise from the outside of the cavity is suppressed by a factor of finesse. 
In addition, any fluctuations of the input laser, including the fluctuations of the polarization, intensity, beam shape, and so on, could be noise sources via practical and spurious couplings. However, the Faraday isolator in the input optical path can suppress the polarization fluctuation. 
The fluctuations of the intensity and beam shape are (or will be) also passively filtered by the input mode cleaner, especially, in the ground-based gravitational wave detector.
Nonetheless, to estimate the sensitivity of each detector, the practical noise should be taken into account by characterizing real detector performances.
% added on 2021/05/12 KN from here
When the noise affects the two ports in a correlated manner, i.e. $C\neq0$, the sensitivity may not be improved by the coherent analysis shown in Section \ref{Two detectors correlation}.
% added on 2021/05/12 KN to here
It is worth noting that the displacement of the cavity can not be a noise source in principle since the displacement does not make polarization rotation ideally.

% \Red{このあたりの記述は全体を見て整える。}

% \section{Sensitivity for for the axion dark matter with the gravitational wave detector}

% \Red{相関解析とかの効果を入れた感度を見せる。}

% \Red{BS入射での偏光雑音とかのコモン系の雑音に関する議論をしたい。}

\section{Conclusion}

In this work, we revisited ADAM-GD scheme with arm cavity transmitted beams and estimated the sensitivity to the axion-photon coupling including the stochastic effect of the axion field oscillation.
We found that the use of transmission port can avoid the cancellation of photon's phase modulation caused by the parity-flipping effect of mirrors and therefore achieves the great sensitivity level for the lower mass range of axion dark matter.
In addition, we analyzed the sensitivity improvement using the two detection ports that is intrinsic to the gravitational wave detectors. Then, we found that the sensitivity is improved by a factor of $\sqrt{2}$ %\SM{\sout{at most} (because it can be more if noise are anti-correlated)} 
with the coherent analysis if the noise of two detection ports are uncorrelated. In the end, the estimated sensitivity of the ground-based detector, such as aLIGO, with 1-year observation can be $g_{a\gamma} \sim 3\times 10^{-12}$ GeV$^{-1}$ below $m<10^{-15}$ eV and potentially overcomes the CAST limit in $m<2 \times 10^{-11}$ eV. We believe that this study leads to the realization of the axion dark matter search with the real gravitational wave detectors.

%====================================================================================%
\section*{Acknowledgement}
%====================================================================================%
\label{Acknowledgement}

We would like to thank Masahiro Ibe for inspiring discussions.
% In this work, KN, YM, TF, HN, SM, and IO are supported by the JSPS KAKENHI Grant  No.~JP20J01928 (KN), No.~JP19J21974 (HN), JSPS Grant-in-Aid for Scientific Research (B) No.~18H01224 (YM), {\it Grant-in-Aid for Transformative Research Areas (A) No.~20H05850 and No.~20H05854 (YM)}, {\it JST PRESTO Grant No.~JPMJPR200B (YM)} and Grant-in-Aid for JSPS Research Fellow No.~17J09103 (TF),
% Advanced Leading Graduate Course for Photon Science (HN),
% and NSF PHY-1912649 (SM), the JSPS Overseas Research Fellowship (IO) and JSPS KAKENHI Grant Number JP20H05859 (IO), respectively.
% \YM{Maybe we don't need our names for all of the grants. Too complicated. How about...}
This work was supported by the JSPS KAKENHI Grant Nos.~JP20J01928, JP19J21974, JP18H01224, JP20H05850, JP20H05854, JP20H05859, JP17J09103, JST PRESTO Grant No.~JPMJPR200B, and NSF PHY-1912649.
KN acknowledges the support from JSPS Research Fellowship.
HN acknowledges the support from the Advanced Leading Graduate Course for Photon Science.
IO acknowledges the support from JSPS Overseas Research Fellowship.

%====================================================================================%
\appendix
\section{probability distribution of $A_N$}
\label{app_derivation_PAN}
%%%%%%%%%%%%%%%%%

The derivation of $P^{(N)}(A_N)$ is explained in this appendix.
The probability distribution of $A^2_N$ is given by
\begin{align}
&\tilde{P}^{(N)}(A^2_N) \nonumber \\
&= \int^{\infty}_0 \left( \prod_i {\rm d} a_i P^{\rm (Ray)}(a_i) \right) \delta\left(A^2_N - \frac{1}{N} \sum_i a^2_i\right) \nonumber \\
&= \int^{\infty}_0 \left( \prod_i {\rm d} b_i \e^{-b_i} \right) \delta\left(A^2_N - \frac{\bar{a}^2}{N} \sum_i b_i\right),
\end{align}
where $b_i \equiv a^2_i / \bar{a}^2$.
After the delta function is replaced by the following integral representation,
\begin{equation}
\delta(x) = \frac{1}{2 \pi} \int^{\infty}_{-\infty} \e^{\iu k x} {\rm d} k,
\end{equation}
the integrations with respect to $b_i$ can be easily performed,
\begin{align}
&\tilde{P}^{(N)}(A^2_N) \nonumber \\
&= \frac{1}{2 \pi} \int^{\infty}_{-\infty} {\rm d} k \e^{\iu k A^2_N} \prod_i \left(\int^{\infty}_0 {\rm d} b_i \e^{-(1 + \iu k \bar{a}^2 / N) b_i} \right) \nonumber \\
&= \frac{1}{2 \pi \iu^N} \left(\frac{N}{\bar{a}^2}\right)^N \int^{\infty}_{-\infty} {\rm d} k \frac{\e^{\iu k A^2_N}}{\left(k - \iu N / \bar{a}^2\right)^N}.
\end{align}
The integrand has a $N$-th order pole at $k=\iu N / \bar{a}^2$, and the integral can be evaluated with the residue theorem,
\begin{equation}
\int^{\infty}_{-\infty} {\rm d} k \frac{\e^{\iu k A^2_N}}{\left(k - \iu N / \bar{a}^2\right)^N} = \frac{2 \pi \iu^N A^{2(N-1)}_N}{\Gamma(N)} \e^{- N \left(A_N / \bar{a}\right)^2}
\end{equation}
Thus, the probability distribution of $A^2_N$ is given by
\begin{equation}
\tilde{P}^{(N)}(A^2_N) = \frac{N^N A^{2(N-1)}_N}{\Gamma(N) \bar{a}^{2N}}  \e^{- N \left(A_N / \bar{a}\right)^2},
\end{equation}
and that of $A_N$ is given by
\begin{align}
P^{(N)}(A_N) &= 2 A_N \tilde{P}^{(N)}(A^2_N) \nonumber \\
&= \frac{2 N^N}{\bar{a} \Gamma(N)} \left(\frac{A_N}{\bar{a}}\right)^{2N - 1} \e^{- N \left(A_N / \bar{a}\right)^2}.
\end{align}

We also comment on the $N$ dependence of the $A^{\mathrm{low}}_{N}(p)$ given by Eq.~\eqref{eq_def_a_alpha}.
In the large $N$ limit, the probability distribution of $A_{N}^2$ becomes Gaussian distribution, and the variance of distribution decreases as $N^{-1/2}$.
As the probability distribution of $A_N$ converges into the sharp peak, the $A^{\mathrm{low}}_{N}(p)$ also converges to the $\bar a$.
Thus, the fitting formula in Eq.~\eqref{eq_fitting_Alow} includes the term proportional to $ N^{-1/2}$ and correction term, which decreases quicker than $N^{-1/2}$ for large $N$. 

% The lower bound of  $A^{\mathrm{low}}_{N}(p)$ is approximately given by
% %by the 1-$\sigma$ and 2-$\sigma$ tail of the probability distribution for $p=0.68$ and $p=0.95$.
% $
%     A^{\mathrm{low}}_{N}(0.68)
%     \to 
%     0.234 \bar a N^{-1/2}
% $
% and 
% $
%     A^{\mathrm{low}}_{N}(0.95)
%     \to 
%     0.822 \bar a N^{-1/2}
% $.

% , where its average and variance is given by
% \begin{align}
%     {\rm E}[A_N^2] 
%     &= N^{-1}\sum_i^N {\rm E}[a_i^2] 
%     = \bar a^2,
%     \\
%     {\rm Var}[A_N^2] 
%     &%={\rm E}[A_N^4] - ({\rm E}[A_N^2])^2
%     = N^{-2}\sum_i^N\sum_j^N {\rm E}[a_i^2a_j^2] - ({\rm E}[A_N^2])^2
%     \\&=
%      N^{-2}\sum_i^N ( {\rm E}[a_i^4] - {\rm E}[a_i^2]^2)
%      =N^{-1}.
% \end{align}

%%%%---------------------------______%%%%%%%%%%%%%%%%%%----------

%\bibliographystyle{apsrev4-2}
%\bibliographystyle{aip}
%\bibliography{library_tex4}

%\if0
% \RED{References will be updated.}

%\fi

\end{document}